**Two-dimensional topological platinum telluride superstructures with periodic tellurium vacancies**

*Xin Xu[1,2,3]†, Xuechun Wang[4]†, Shuming Yu[5]†, Chenhui Wang[3,2], Guowei Liu[6], Hao Li[2], Jiangang Yang[2], Jing Li[7], Tao Sun[8], Xiao Hai[9], Lei Li[10], Xue Liu[11], Ying Zhang[10], Weifeng Zhang[3,12], Quan Zhang[13], Kedong Wang[6], Nan Xu[5], Yaping Ma[3,2,12]\*, Fangfei Ming[1]\*, Ping Cui[4]\*, Jiong Lu[9], Zhenyu Zhang[4], Xudong Xiao[2]\**

[1]State Key Lab of Optoelectronic Materials and Technologies, Guangdong Province Key Laboratory of Display Material and Technology and School of Electronics and Information Technology, Sun Yat-sen University, Guangzhou 510275, People's Republic of China

[2]School of Physical Science and Technology and Key Laboratory of Artificial Micro- and Nano-Structures of Ministry of Education, Wuhan University, Wuhan 430072, People's Republic of China

[3]School of Future Technology, Henan Key Laboratory of Quantum Materials and Quantum Energy, Henan University, Zhengzhou 450046, People's Republic of China

[4]International Center for Quantum Design of Functional Materials (ICQD), University of Science and Technology of China, Hefei 230026, People's Republic of China

[5]Institute for Advanced Studies, Wuhan University, Wuhan 430072, China

[6]Department of Physics, Southern University of Science and Technology, Shenzhen 518055, People's Republic of China

[7]Key Laboratory of Bio-inspired Smart Interfacial Science and Technology of Ministry of Education, School of Chemistry, Beihang University, Beijing 100191, China

[8]School of Chemical Engineering, Xi'an Key Laboratory of Special Energy Materials, Northwest University, Xi'an 710069, People's Republic of China

[9]Department of Chemistry, National University of Singapore, Singapore 117543, Singapore

[10]Core Facility of Wuhan University, Wuhan 430072, China

[11]Anhui Key Laboratory of Magnetic Functional Materials and Devices, Institute of Physical Science and Information Technology, Anhui University, Hefei, Anhui, 230601 China

[12]Institute of Quantum Materials and Physics, Henan Academy of Sciences, Zhengzhou 450046, People's Republic of China

[13]College of Chemistry and Chemical Engineering, Hubei Normal University, Huangshi 435002, People's Republic of China

**ABSTRACT**

Defect engineering in the inherently inert basal planes of transition metal dichalcogenides (TMDs), involving the introduction of chalcogen vacancies, represents a pivotal approach to enhance catalytic activity by exposing high-density catalytic metal single-atom sites. However, achieving a single-atom limit spacing between chalcogen vacancies to form ordered superstructures remains challenging for creating uniformly distributed high-density metal single-atom sites on TMDs comparable to carbon-supported single-atom catalysts (SACs). Here we unveil an efficient TMD-based topological catalyst for hydrogen evolution reaction (HER), featuring high-density single-atom reactive centers on a few-layer (7×7)-$PtTe_{2-x}$ superstructure. Compared with pristine Pt(111), $PtTe_2$, and (2×2)-$PtTe_{2-x}$, (7×7)-$PtTe_{2-x}$ exhibits superior HER performance owing to its substantially increased density of undercoordinated Pt sites, alongside exceptional catalytic stability when operating at high current densities. First-principles calculations confirm that multiple types of undercoordinated Pt sites on (7×7)-$PtTe_{2-x}$ exhibit favorable hydrogen adsorption Gibbs free energies, and remain active upon increasing hydrogen coverage. Furthermore, (7×7)-$PtTe_{2-x}$ possesses nontrivial band topologies with robust edge states, suggesting potential enhancements for HER. Our findings open up a new avenue for the development of TMD-based catalysts and exciting prospects for the exploration of topological materials in catalysis and beyond.

**INTRODUCTION**

Developing a carbon-neutral economy requires replacing our reliance on fossil fuels with sustainable fuels and chemical building blocks from renewable resources. Utilizing advanced materials and environmentally friendly technologies such as electrocatalysis to convert renewable resources (e.g., $H_2O$, $CO_2$, $N_2$, and solar energy) into value-added products (e.g., $H_2$, hydrocarbons, and $NH_3$) provides an effective approach to address these problems.[1,2] For instance, two-dimensional (2D) transition metal dichalcogenides (TMDs) have been proven to be promising candidates for hydrogen production *via* electrocatalytic water splitting.[3-7] Defect engineering in the intrinsically inert basal planes of TMDs (e.g., vacancies, boundaries, heteroatom doping, etc.) is an effective strategy for introducing the catalytically active sites.[5-10] Apart from introducing catalytically active defective sites, achieving high-density of intrinsic catalytic sites holds conceptually superior significance. Typically, chalcogen vacancies are introduced to expose transition metal atom sites for hydrogen adsorption. The density of metal atom sites is strongly correlated with the average spacing between chalcogen vacancies. By reducing this average spacing close to the single-atom limit while maintaining the activity of exposed metal sites, the density of metal atom sites can be significantly increased.[11] If the spacing between each two chalcogen vacancies can be reduced to the single-atom limit like atomic Lego, forming an ordered superstructure, a high-density of metal single-atom sites will be exposed (Figure 1a). Nevertheless, the creation of high-density active single-atom sites on TMDs, comparable to the dense active sites in single-atom catalysts (SACs),[12,13] remains a formidable challenge due to the unpredictable formation of various defects with diversely varying catalytic activities.

As a compelling, platinum dichalcogenides belong to the 2D noble TMD family with the van der Waals-type interlayer interactions.[14,15] These materials have drawn great attention recently due to their excellent environmental stability, high carrier mobility, and layer-dependent band structures, offering appealing opportunities for applications in nanoelectronics, optoelectronics and catalysis.[15-21] However, in the pristine structure, Pt atoms are hexacoordinated and enclosed between the top and bottom chalcogen atom layers, resulting in a chemically inert basal plane.[22] Various approaches such as heat treatment, and $Ar^+$ sputtering have been employed to expose abundant active defect sites in $PtTe_2$ to enhance its electrocatalytic activity for hydrogen evolution reaction (HER).[5,23,24] Despite many recent advances, controllable strategies of structural engineering for creating high-density catalytic sites still pose standing challenges in the design and fabrication of highly efficient TMD-based electrocatalysts.

Here we report a novel single-atom topological catalyst of $PtTe_{2-x}$ with a high density of catalytically active Pt sites (> 5.0 sites nm$^{-2}$) for HER electrocatalysis, by the precise bottom-up nanofabrication of few-layer $PtTe_{2-x}$ films to form well-ordered Te-vacancy superstructures. Our approach involves a stepwise annealing process: a few-layer Te film grown on a single-crystal Pt(111) substrate is first annealed at low temperatures, transforming it into a $PtTe_2$ film. Further annealing at elevated temperatures results in the induction of Te deficiency, yielding well-ordered (2×2)- and (7×7)-$PtTe_{2-x}$ superstructures revealed by high-resolution scanning tunneling microscopy (STM) and transmission electron microscopy (TEM) coupled with the simulations of density functional theory (DFT) calculations. Scanning tunneling spectra (STS) and angle-resolved photoemission spectroscopy (ARPES) measurements demonstrate the metallic nature of both structures. The Te-vacancy-induced undercoordinated Pt sites in the atomically flat (7×7)-PtTe$_{2-x}$ superstructure exhibit excellent electrocatalytic activity, experimentally showing an overpotential of 42 mV at 10 mA cm$^{-2}$ and a Tafel slope of 56.7 mV dec$^{-1}$, surpassing the HER performance of PtTe$_2$ and (2×2)-PtTe$_{2-x}$, and even better than that of pristine Pt(111). Our first-principles calculations demonstrate that there are eight types of undercoordinated Pt sites in the (7×7)-PtTe$_{2-x}$ structure, and all of them exhibit excellent hydrogen adsorption Gibbs free energies ($\Delta G_{H*}$) in the low H coverage regime, with the lowest $\Delta G_{H*}$ being -7 meV. Meanwhile, the undercoordinated Pt sites continue to be energetically active for HER with increasing hydrogen coverage. Moreover, the (7×7)-PtTe$_{2-x}$ superstructure possesses nontrivial band topologies, and the accompanying robust edge states may further facilitate HER.

**RESULTS AND DISCUSSION**

**Synthesis of PtTe$_{2-x}$ superstructures.** The sample fabrication strategy used in our experiment is depicted in Figure 1b. An atomically clean Pt(111) surface serves as the substrate, whose STM image shows the presence of wide terraces and single atomic steps (Figure S1a). Upon deposition of tellurium, we observe the formation of atomically flat films with step heights of ~ 4.0 Å (Figure S1b). The small-scale STM image (Figure S1c) reveals a rectangular lattice with lattice constants of $a$ =4.43 Å and $b$ =5.95 Å. Moreover, STS measurements (Figure S1d) unveil a semiconducting gap of ~0.83 eV. These observations are in accordance with the characteristics of the β-phase tellurene (β-Te) typically found when grown on graphene or HOPG substrates.[25,26] To the best of our knowledge, this work represents the first successful report of the fabrication of crystalline β-Te films on single crystalline metal substrates.[27-29]

Subsequently, few-layer Pt-Te compound films are prepared by post-annealing the β-Te film on Pt(111). With the elevation of the annealing temperature, the surface structures undergo an evolution from a hexagonal 1T-PtTe$_2$ film at 150 °C (Figure 1c,d), to a (2×2)-PtTe$_{2-x}$ superstructure at 250 °C (Figure 1e,f), and finally to a (7×7)-PtTe$_{2-x}$ superstructure with ordered and well-defined Te vacancies at 400 °C (Figure 1g,h). Here, x denotes the presence and density of the Te vacancies. The results of the area fractions of different phases with different sample annealing conditions are summarized in Figure S2a. By carefully controlling the annealing conditions, we can achieve complete surface coverage of the (7×7)-PtTe$_{2-x}$ superstructures (Figure S2b). After annealing at 150 °C, 250 °C, and 400 °C, PtTe$_2$, (2×2)-PtTe$_{2-x}$ and (7×7)-PtTe$_{2-x}$ are dominating, respectively, while several other intermediate phases are also observed under intermediate annealing conditions (Figure S3). Since high-temperature annealing is well known to induce Te deficiencies, all the phases other than PtTe$_2$ are expected to be Te-deficient Pt-Te compounds. Indeed, after depositing extra Te atoms on the PtTe$_{2-x}$ structures and annealing at 150 °C, the surface recovers to pure PtTe$_2$ (Figure S3).

More specifically, upon annealing at 150 °C for 30 min, the β-Te structure vanishes and the surface is fully covered with the PtTe$_2$ film (Figure 1c,d and Figure S4a). The resulting surface exhibits flat terraces with hexagonally shaped plateaus, featuring atomic step heights of ~5.0 Å. Aside from a few disordered regions, the surface largely manifests a triangular lattice with a lattice constant of ~4.1 Å (Figure S4b), which is in excellent agreement with the characteristics of 1T-phase PtTe$_2$.[30,31] Our calculated STM image using 1T-PtTe$_2$ (Figure S4c) also shows a hexagonal lattice of the top-layer Te atoms, consistent with the experimental image. After further annealing at 250 °C for 30 minutes, the hexagonal plateaus disappear, and the terrace widths become wider, while some disordered areas still exist (Figure S4d). High-resolution STM images (Figure 1f and Figure S4e) show the presence of a distinct Kagome-type lattice consisting of a (2×2) quadruple of the PtTe$_2$ lattice unit cell. Notably, the depressions in every (2×2) superstructure of the 1T-PtTe$_2$ surface, as well as the low-energy electron diffraction (LEED) pattern (Figure 1e), provide evidence for the formation of ordered Te vacancies, which are spaced at the single-atom limit. This particular superstructure is denoted as (2×2)-PtTe$_{2-x}$. The initial atomic model of (2×2)-PtTe$_{2-x}$ is constructed based on the PtTe$_2$ structure, involving the removal of one-quarter of the Te atoms from the top Te layer, as guided by atomically resolved STM imaging. Subsequent structural relaxation yields a calculated STM image (Figure S4f) that closely matches the experimental results. Moreover, the stability of the (2×2)-PtTe$_{2-x}$ superstructure (PtTe$_{1.75}$) has been verified through theoretical calculations of formation energy, phonon dispersion, and molecular dynamics simulations.[21,32,33] After undergoing additional annealing at 400 °C for 30 min, the surface attains greater uniformity, featuring flat terraces with the emergence of small nanometer-sized islands at the step edges (Figure S4g). To our surprise, the atomic resolution STM image and LEED pattern reveal the presence of a (7×7) superstructure relative to the $PtTe_2$ lattice, denoted as (7×7)-$PtTe_{2-x}$ (Figure 1h and Figure S4h). This structure appears to be comprised of identical (2×2)-$PtTe_{2-x}$ domain patches separated by domain boundaries. One of these domain patches is labeled with a blue Star of David (Figure S4g), while the domain boundary between three nearest neighboring domain patches shows a bright trimer structure, as labeled by a red triangle. Overall, this newly discovered structure presents a striking Kagome-like lattice pattern with a (7×7) periodicity, as indicated by the red Star of David shape. To rule out the possibility that the superstructures of (2×2)- and (7×7)-$PtTe_{2-x}$ stem from charge density wave (CDW) phases,[34] we verified the consistency of the STM images obtained at room temperature with those acquired at low temperature (Figure S5).

To further confirm that our structures are indeed formed on top of few-layered $PtTe_{2-x}$, we have conducted transmission electron microscopic (TEM) imaging on the cross-section of the $PtTe_2$ and (7×7)-$PtTe_{2-x}$ samples. High-resolution TEM and aberration-corrected high-angle annular dark field scanning TEM (HAADF-STEM) images (Figure 2a) reveal the formation of multilayer $PtTe_2$ with a thickness of ~6.6 nm, displaying a Te-Pt-Te sandwich configuration in each layer. This configuration aligns with the coexistence of Pt and Te elements as confirmed by energy-dispersive X-ray spectroscopy (EDS) mapping (Figure S6). These results unambiguously demonstrate the formation of multilayer $PtTe_2$ on Pt(111) surface via direct tellurization. For the (7×7)-$PtTe_{2-x}$ sample, the large-scale TEM image shows a thickness reduction to 2-5 nm, indicating Te loss during annealing (Figure 2b). High-resolution TEM image (Figure 2c) displays a quadra-layer structure of (7×7)-$PtTe_{2-x}$ with a lattice spacing of ~5.3 Å, corresponding to the (001) plane of the $PtTe_2$ crystal (inset of Figure 2c). The HAADF-STEM image further reveals this quadra-layer Te-Pt-Te sandwich structure (layer-to-layer spacing: 5.21 Å) near the (7×7)-$PtTe_{2-x}$/Pt(111) interface, consistent with the Pt and Te elements identified in EDS mapping (Figure 2d). Additionally, atomic-resolution HAADF-STEM image at the Te protective layer/(7×7)-$PtTe_{2-x}$ interface (Figure 2e) shows a distinct intensity reduction in the top Te atomic columns of the (7×7)-$PtTe_{2-x}$ surface layer, indicating Te deficiency, while the underlying layers show no significant intensity difference between the top and bottom Te layers, implying the preservation of the PtTe$_2$ phase. The atomic ratios of Pt and Te elements in a quadra-layer (7×7)-PtTe$_{2-x}$ thin film are 35.90% and 64.10%, respectively, corresponding to a Pt/Te ratio of ~0.56 (Figure S7), further supporting the PtTe$_{2-x}$ structures. Therefore, based on the top-view STM and side-view TEM images, we constructed an atomic model of (7×7)-PtTe$_{2-x}$ (Pt$_{49}$Te$_{85}$) by arranging small patches of (2×2)-PtTe$_{2-x}$ following the atomic-resolution STM image, while the bottom layer Te atoms are kept intact. These identically sized (2×2)-PtTe$_{2-x}$ patches are deliberately shifted out of phase, giving rise to modified Te-vacancy configurations at the domain boundaries. Upon structural relaxation (Figure 2f), two types of boundary configurations are formed: one with a threefold undercoordinated Pt atom surrounded by six onefold undercoordinated Pt atoms, and another with three onefold undercoordinated and three twofold undercoordinated Pt atoms arranged in a triangular shape. These two boundaries produce similar bright trimer structures in the simulated STM images (bottom of Figure 2f and Figure S4i), showing excellent agreement with the experimental STM results. Our STM images shown above should be compared with recently reported 2D surface telluride phases on Pt(111) under relatively low Te dosage.[35] Surprisingly, our prepared (2×2) and (7×7) superstructures with respect to the PtTe$_2$ lattice exhibit striking similarities respectively to their (3×3) and (10×10) superstructures with respect to the Pt(111) lattice. However, as demonstrated by the above TEM images, despite the similar surface features revealed in STM images, our samples comprise of few-layer PtTe$_{2-x}$, in contrast to the reported surface telluride phases.

Our results above further indicate that PtTe$_{2-x}$ phases can be synthesized through tellurization of Pt(111), aligning with the formation mechanism of PtSe$_2$ via direct selenization[36-38] and a prior report of (2×2)-PtTe$_{2-x}$ formation by annealing the PtTe$_2$/Pt(111) system.[23] Our discovery of the (7×7)-PtTe$_{2-x}$ structure in particular is beyond existing literature. In the past, due to the importance of atomic defects in TMDs, such as chalcogen vacancies and substitutions, in enhancing the performance of electronic/optoelectronic devices and catalysis,[39] various techniques, including ion plasma, laser irradiation, electron beam exposure, and thermal annealing, have been employed to introduce chalcogen atomic defects in TMDs.[40-43] Unfortunately, these methods typically produce a random distribution of various types of chalcogen atomic defects. For instance, thermal treatment has been reported to drive the migration of Te single vacancies in PtTe$_2$, leading to the formation of Te-vacancy trimer clusters, which randomly appear on the PtTe$_2$ surface.[5] Even using the same growth procedures on a Pt polycrystalline thin film on a silicon substrate, the preparation of PtTe$_2$ phase was unsuccessful (Figure S8). This failure is likely attributed to the inability of the Pt polycrystalline thin film to undergo tellurization at the low temperature of 150 °C. Our successful preparation of ordered PtTe$_{2-x}$ phases on crystalline Pt(111) substrate is one of the few examples that controllable creation of periodic chalcogen-vacancy superstructures uniformly over a large-area has been achieved. This development is essential for quantitative understanding the mechanisms of atomic defects in TMDs for various applications.

**Characteristics of PtTe$_{2-x}$ films.** To investigate the characteristics of various Pt-Te compound films on the Pt(111) substrate, we conducted Raman, *in-situ* X-ray photoemission spectroscopy (XPS), and ARPES measurements. The Raman spectrum of the as-grown PtTe$_2$ (Figure 3a) reveals the presence of two primary peaks at ~120 and 156 cm$^{-1}$, corresponding to the in-plane phonon mode E$_g$ and the out-of-plane phonon mode A$_g$ of PtTe$_2$, respectively.[16,44,45] Additionally, a minor peak at approximately 94 cm$^{-1}$, attributed to the E$_g$ mode of PtTe,[45] is observed due to a slight Te deficiency. Upon elevating the annealing temperature to facilitate the creation of (2×2)- and (7×7)-PtTe$_{2-x}$ phases with more Te deficiencies, we observed a discernible enhancement in the peak at ~94 cm$^{-1}$. In particular, two new peaks emerge at ~169 and 184 cm$^{-1}$ for (7×7)-PtTe$_{2-x}$, corresponding to the A$_{1g}$ and E$_g$ modes of PtTe, respectively.[45] The chemical states of Pt and Te in various PtTe$_{2-x}$ phases are characterized by XPS (Figure 3b and Note S6). The spin-orbit-split Pt-4f$_{5/2}$ and Pt-4f$_{7/2}$ peaks of the PtTe$_2$ film are situated at ~75.6 and 72.3 eV respectively (corresponding to Pt$^{4+}$), congruent with that of the previously reported platinum ditelluride (PtTe$_2$).[18] Notably, the absence of Pt$^0$ signals that emanate from the underlying Pt substrate reveals a discernible thickness of the PtTe$_2$ film (Figure S9), supported by the observation of thickness exceeding 6 nm from TEM image of this multilayer PtTe$_2$ (Figure S6). For the (2×2)- and (7×7)-PtTe$_{2-x}$ phases, alongside the Pt$^{4+}$ peaks, an additional pair of peaks manifests at ~74.7 and 71.4 eV, matching well with the Pt-4f peaks of Pt$^{2+}$ in platinum telluride (PtTe),[45,46] as a consequence of Te deficiency. Moreover, the detection of the Pt-4f peaks of Pt$^0$, at ~74.3 and 71.0 eV, signifies that the acquired (2×2)- and (7×7)-PtTe$_{2-x}$ films are thinner in nature compared to the PtTe$_2$ film, consistent with the TEM results. The line shape and peak positions of Te-3d of the PtTe$_2$ and PtTe$_{2-x}$ films, located at ~583.5 and 573.1 eV, remain predominantly unaltered (Figure S9).[45] These distinctive findings diverge from prior reports concerning 2D Te,[20] affirming the formation of Pt-Te compound films on the Pt(111) surface, rather than the presence of 2D Te or Te reconstruction.

The electronic properties of the PtTe$_2$, (2×2)- and (7×7)-PtTe$_{2-x}$ films are probed through STS and ARPES. The typical STS spectra of the PtTe$_2$, (2×2)- and (7×7)-PtTe$_{2-x}$ films consistently exhibit metallic behaviors (Figure S11a). It is worth noting that the metallic characteristics observed in PtTe$_2$ indicate the presence of a multilayered structure (Figure S11b-d). The ARPES results for the pristine Pt(111) surface are summarized in Figure S12, and the energy-wave vector dispersions along two high-symmetry ΓK and ΓM directions of the surface Brillouin zone (BZ) exhibit both electron and hole pockets at the binding energy of ~1.5 eV, consistent with previous results.[47] Regarding the PtTe$_2$ film, the constant energy (CE) intensity map at the Fermi energy (Figure S13b) exhibits discernible features, confirming its metallic nature and multi-layer Te-Pt-Te sandwich structure. Moreover, the dispersions (Figure S13d,e) are in congruence with that of multilayer PtTe$_2$ structure.[48] In the case of (2×2)-PtTe$_{2-x}$ film, the dispersion images (Figure S14c,d) unveil several weak bands below the binding energy of 1.5 eV. It is noted that the visible Pt(111) surface band (comparing Figure S14 with Figure S12) indicates a reduced thickness for the (2×2)-PtTe$_{2-x}$ film compared to the PtTe$_2$ film. The (7×7)-PtTe$_{2-x}$ film also exhibits a metallic nature, as evidenced by the CE intensity map at the Fermi energy (Figure S15a). The dispersion images (Figure 2c,e and Figure S15c,d) illustrate a pronounced hole pocket at the binding energy of about 0.13 eV. Furthermore, the second-order derivative of the dispersion images (Figure 2d,f) reveals three faint bands at binding energies of ~0.57, 0.76, and 0.94 eV, which are largely consistent with the calculated band structures. Again, the observation of Pt(111) states in the dispersion images confirms the decreased thickness of the (7×7)-PtTe$_{2-x}$ film in comparison to the PtTe$_2$ film.

**HER performance of PtTe$_{2-x}$ catalysts.** Both the (2×2)- and (7×7)-PtTe$_{2-x}$ structures expose substantial densities of undercoordinated Pt sites (~ 5.28 and 5.08 sites nm$^{-2}$ for (2×2)- and (7×7)-PtTe$_{2-x}$, respectively, as calculated in Note S7), which are comparable to the densities of active atoms in SACs,[13] and are beneficial to electrocatalysis. Air stability of the PtTe$_{2-x}$ structures is a prerequisite for ambient applications. As anticipated, the PtTe$_2$ structure exhibits exceptionally high stability under ambient conditions, due to its perfect lattice plane (Figure S16). After air exposure, the overall structures, including the terraces and step edges, of the (2×2)- and (7×7)-PtTe$_{2-x}$ surfaces remain essentially unchanged, although these surfaces appear to be mildly contaminated by an atomically thin adsorbate layer, presumed to be primarily diffusing gas molecules (Figure S17a,b). Subsequent annealing at 150 °C can effectively remove those contaminants, restoring most surface areas to their well-ordered structures (Figure S17c,d), a fact further substantiated by LEED patterns (insets of Figure S17 and Figure S18). In addition, Auger electron spectroscopy (AES) measurements indicate minimal oxidation during air exposure (Figure S17e,f). Collectively, these experiments provide compelling evidences for the air stability of the (2×2)- and (7×7)-PtTe$_{2-x}$ surfaces.

Hereafter, we perform HER test experiments with an acidic electrolyte (0.5 M H$_2$SO$_4$) to evaluate the electrocatalytic performance of the PtTe$_2$, (2×2)- and (7×7)-PtTe$_{2-x}$ samples. The experimental setup is illustrated in Figure S19, and the tested samples include the PtTe$_2$ and PtTe$_{2-x}$, together with the clean Pt(111) surface and the state-of-the-art 20 wt.% Pt/C catalyst (Figure S20) for comparison. The linear sweep voltammetry (LSV) curves (Figure 4a) indicate that all the samples exhibit significant HER current densities ($J$) except the Ta substrate, showing the minimal influence of the Ta substrate on the HER performance of the tested samples. HER activity is gradually enhanced from PtTe$_2$, to (2×2)-PtTe$_{2-x}$ and to (7×7)-PtTe$_{2-x}$, as evidenced by their decreasing Tafel slopes and overpotentials at a current density of 10 mA cm$^{-2}$ (Figure 4b and Figure S21a). Here, the Tafel slope (56.7 mV dec$^{-1}$) of (7×7)-PtTe$_{2-x}$ is even lower than that of Pt(111), although it is still larger than that of the Pt/C catalyst (33.0 mV dec$^{-1}$). The (7×7)-PtTe$_{2-x}$ sample only requires an overpotential ($\eta$) of 42 mV at a current density of 10 mA cm$^{-2}$ (Figure 4b and Figure S21b), which is much lower than that of PtTe$_2$ (168 mV), (2×2)-PtTe$_{2-x}$ (140 mV), and Pt(111) (120 mV), while comparable to that of Pt/C (36 mV). The exchange current density ($J_0$) of (7×7)-PtTe$_{2-x}$ (1.90 mA cm$^{-2}$), obtained by extrapolating the Tafel plots (Figure S21a, similar to previously reported method[49]), outperforms those of PtTe$_2$ (0.97 mA cm$^{-2}$), (2×2)-PtTe$_{2-x}$ (0.23 mA cm$^{-2}$), and Pt(111) (0.30 mA cm$^{-2}$), indicating that the (7×7)-PtTe$_{2-x}$ structure has the highest inherent catalytic activity. Due to the complex kinetic processes in HER, the exchange current density and Tafel slope are not entirely independent. HER electrocatalysts with lower Tafel slopes tend to exhibit smaller exchange current densities, and *vice versa*.[50] Hence, a smaller overpotential at the targeted current density is always the figure of merit for evaluating a better electrocatalyst.[50] In addition, among these samples, (7×7)-PtTe$_{2-x}$ delivers the highest turnover frequency (TOF) value (~452.0 s$^{-1}$) at the overpotential of 100 mV, surpassing significantly the reported TOF values of single-atom catalysts (Figure S21c,d).[51-56]

Moreover, electrochemical surface area (ECSA), as a key indicator to evaluate the intrinsic catalytic activity can be determined by the double layer capacitance ($C_{dl}$), which is obtained from cyclic voltammetry (CV) curves at various scan rates within the non-Faradic region (Figure S22).[57] The (7×7)-PtTe$_{2-x}$ sample exhibits the largest $C_{dl}$ (19.2 mF cm$^{-2}$), indicating that it has more active sites for HER than other Pt-based structures (Figure S23a). As the rough surface of the Pt/C catalyst exposes more Pt sites, leading to a much larger ECSA (Figure S24), the HER performance of the Pt/C catalyst is still better than that of the (7×7)-PtTe$_{2-x}$ sample. To evaluate the intrinsic HER activities of the catalytic sites, the electrocatalytic current densities are normalized by $C_{dl}$, denoted as $J_{spc}$. Compared to PtTe$_2$, (2×2)-PtTe$_{2-x}$, and Pt(111), the undercoordinated Pt sites of (7×7)-PtTe$_{2-x}$ possess superior intrinsic activities (Figure 4c).[58] Notably, these intrinsic activities are marginally superior to the Pt/C catalyst. The electrochemical impedance spectroscopy (EIS) measurements (Figure S23b) demonstrate that the charge transfer resistance of the (7×7)-PtTe$_{2-x}$ sample is significantly lower than that of PtTe$_2$, (2×2)-PtTe$_{2-x}$ and Pt(111), again indicating its superior HER activity.

In addition to assessing the catalytic activity, the catalytic stability is of paramount importance. The long-term chronopotentiometry measurement (Figure 4d) shows extremely high stability for over 90 h with a minimal change in overpotential at a constant current density of 100 mA cm$^{-2}$ for the (7×7)-PtTe$_{2-x}$ sample, which significantly surpasses that of the Pt/C catalyst. Furthermore, the (7×7)-PtTe$_{2-x}$ sample exhibit robust long-term stability for over 70 h even at a substantial current density of 300 mA cm$^{-2}$, bolstering its potential for industrial-scale applications. In the accelerated CV test, the (7×7)-PtTe$_{2-x}$ sample demonstrates robustness in HER catalysis, with nearly unchanged polarization curves even after 10,000 CV cycles (Figure S23c), while the Pt/C catalyst is less stable during such a test. Moreover, the STM image, LEED pattern, XPS spectra and TEM characterizations of the (7×7)-PtTe$_{2-x}$ sample after the HER test reveal the structural integrity and stability of (7×7)-PtTe$_{2-x}$ during HER operation (Figure S23d-h). Finally, the HER performance of (7×7)-PtTe$_{2-x}$ in an alkaline electrolyte was also tested. Compared with Pt(111), the (7×7)-PtTe$_{2-x}$ sample exhibits better HER performance with an overpotential of 82 mV at a current density of 10 mA cm$^{-2}$ and a Tafel slope of 123.2 mV dec$^{-1}$ (Figure S23i,j). In addition, the (7×7)-PtTe$_{2-x}$ sample demonstrates excellent structural stability in an alkaline electrolyte (Figure S23k).

Compared to similar TMD-based catalysts in an acidic environment, the (7×7)-PtTe$_{2-x}$ structure exhibits outstanding HER activity with the lowest overpotential at the current density of 10 mA cm$^{-2}$ and the moderate Tafel slope (Figure 4e and Table S1).[3,10,11,59-62] Thermally induced Te-vacancy trimer clusters in PtTe$_2$ have been proven to significantly enhance the HER performance, achieving an overpotential of 22 mV at 10 mA cm$^{-2}$ and aTafel slope of 29.9 mV dec$^{-1}$, surpassing the performance of the Pt/C catalyst.[5] Unlike the PtTe$_2$ nanoflakes decorated with randomly dispersed Te-vacancy trimer clusters, we have realized two distinct PtTe$_{2-x}$ superstructures, (2×2) and (7×7), featuring ordered Te-vacancy lattices. Comprehensive characterizations using STM, LEED, ARPES, XPS, and TEM have confirmed the ordered nature of Te vacancies, allowing us to estimate the densities of undercoordinated Pt sites as ~5.28 and 5.08 sites nm$^{-2}$ for the (2×2) and (7×7) superstructures, respectively, comparable to the density of metal atoms in SACs. The exposed Pt sites in the (7×7)-PtTe$_{2-x}$ superstructure exhibit superior intrinsic activities, as corroborated by the TOF value and theoretical results discussed in the following section. Additionally, the ultrathin flat (7×7)-PtTe$_{2-x}$ film acts as a self-supporting electrode with a minimal loading coverage of ~1.84 μg cm$^{-2}$ for a quadra-layer model, substantially lower than that that required for active loading materials on glassy carbon (~0.1-1 mg cm$^{-2}$). This low loading amount may contribute to the observed lower HER performance of the flat (7×7)-PtTe$_{2-x}$ film compared to PtTe$_2$ nanoflakes with Te-vacancy trimer clusters. Nonetheless, the mass activity of (7×7)-PtTe$_{2-x}$ demonstrates an enhancement over the commercial Pt/C catalyst (Figure 4f) by three orders of magnitude. The significantly reduced Tafel slope of (7×7)-PtTe$_{2-x}$ compared with pristine

PtTe$_2$ highlights the advantageous impact of ordered Te vacancies on electrocatalytic HER.[63,64] While the recovery of noble metal Pt in PtTe$_{2-x}$ catalysts prepared on Pt(111) surface presents significant challenges, it is important to note that the recovery of Pt from electrochemically exfoliated PtTe$_2$ nanoflakes can be effectively achieved using various established methods.[65]

**Theoretical insights and topological catalysis.** To elucidate the origin of the HER activities in the PtTe$_{2-x}$ samples, we systematically investigate the hydrogen adsorption Gibbs free energies ($\Delta G_{H^*}$, a metric for assessing HER performance[66]) at different surface sites using DFT calculations. For HER, the binding between a hydrogen adatom and a catalyst should be neither too strong nor too weak for the reaction to proceed, yielding an optimal $\Delta G_{H^*}$ close to zero.[67] As reference systems, the active Pt(111) and inert PtTe$_2$ surfaces are also considered. $\Delta G_{H^*}$ on Pt(111) is -0.095 eV, confirming its excellent HER activity, while the lowest $\Delta G_{H^*}$ on PtTe$_2$ is 1.056 eV, indicating that the hydrogen atom only can be weakly adsorbed on PtTe$_2$ and that the Volmer adsorption process will be hindered. Indeed, the hydrogen atom prefers to stay at a bridge site between two surface Te atoms rather than be directly adsorbed on middle Pt atoms, resulting in weak binding. In contrast, both the (2×2)-PtTe$_{2-x}$ and (7×7)-PtTe$_{2-x}$ structures expose several Pt sites on their surfaces due to Te vacancies, which should significantly enhance the hydrogen binding, as demonstrated in TMDs with defects for HER,[68,69] and also confirmed by our DFT calculations below.

For the (2×2)-PtTe$_{2-x}$ surface, there is only one type of undercoordinated Pt site (Figure S25), in which a Pt atom bonds with five Te atoms (note that the coordination number of Pt in PtTe$_2$ is six) and serves as the active site for hydrogen adsorption. The calculated $\Delta G_{H^*}$ is -0.119 eV, indicating excellent HER activity of (2×2)-PtTe$_{2-x}$. The (7×7)-PtTe$_{2-x}$ structure with more Te vacancies possesses eight types of undercoordinated Pt sites, labelled as Pt1 to Pt8 (Figure S26). Each Pt atom in the Pt2 to Pt7 types bonds with five Te atoms, similar to the undercoordinated Pt site in (2×2)-PtTe$_{2-x}$, but has slightly weaker hydrogen binding. The resultant $\Delta G_{H^*}$ values (Figure 5a) are all positive but much closer to zero than that of (2×2)-PtTe$_{2-x}$, and the values on Pt2 to Pt6 are even closer to zero than that of Pt(111). Moreover, Pt1 and Pt8, with three and two undercoordination numbers, respectively, are still active for HER due to their favorable $\Delta G_{H^*}$ values. A comparison of $\Delta G_{H^*}$ for the Pt(111), PtTe$_2$, (2×2)-

PtTe$_{2-x}$, and (7×7)-PtTe$_{2-x}$ (using the value of -0.007 eV on Pt8) surfaces is displayed in Figure 5b (see Figure S25 for the corresponding atomic models). Notably, ΔG$_{H*}$ of both (2×2)-PtTe$_{2-x}$ and (7×7)-PtTe$_{2-x}$ outperforms that of PtTe$_2$ by a considerable margin. In particular, as shown in Figure 5c, the (7×7)-PtTe$_{2-x}$ structure approaches the volcanic peak with $J_0$ =1.9 mA cm$^{-2}$, even better than that of the commercial Pt/C catalyst.

To examine the influence of hydrogen coverage ($\theta$) on the catalytic activity for HER, we calculate ΔG$_{H*}$ at different hydrogen coverages for both (2×2)- and (7×7)-PtTe$_{2-x}$. Here, the hydrogen coverage is defined as the ratio of the number of hydrogen adatom(s) to the total number of undercoordinated Pt sites on the surface. For (2×2)-PtTe$_{2-x}$, ΔG$_{H*}$ changes little at low coverages (e.g., $\theta$ < 1/3), even if two or three hydrogen atoms are adsorbed on the nearest neighboring Pt sites around a Te vacancy, while it increases slightly towards the zero value with further increasing $\theta$, and reaches -0.056 eV at $\theta$ = 1 (Figure S27). For (7×7)-PtTe$_{2-x}$, the nearest neighboring hydrogen adsorption has a pronounced desirable impact on ΔG$_{H*}$, particularly involving the Pt1 and Pt8 sites (Tables S2 and S3). When two hydrogen atoms are adsorbed on two nearest neighboring sites from Pt1 to Pt8, ΔG$_{H*}$ only exhibits small changes of less than 50 meV (Figure 5d and Table S3). In addition, the change in ΔG$_{H*}$ remains minimal at different hydrogen coverages for (7×7)-PtTe$_{2-x}$ (Figure 5e). Collectively, with increasing $\theta$, both (2×2)-PtTe$_{2-x}$ and (7×7)-PtTe$_{2-x}$ can continue to serve as effective catalysts for HER with ample active sites. Moreover, the solvent effect on the ΔG$_{H*}$ of the PtTe$_{2-x}$ structures is found to be negligible, varying within 10 meV (Table S4).

To gain deeper mechanistic insights into the favorable ΔG$_{H*}$ on the Pt sites around Te vacancies, the charge density difference and charge transfer have been analyzed.[70] Our DFT calculations show that negligible charge transfer occurs for hydrogen adsorption on PtTe$_2$ due to weak binding, while effective charge transfer from the hydrogen to the undercoordinated Pt sites is observed on (2×2)-PtTe$_{2-x}$ and (7×7)-PtTe$_{2-x}$, leading to stronger hydrogen binding and an optimal ΔG$_{H*}$ for HER (see, for example, charge transfer on (7×7)-PtTe$_{2-x}$ in Figure S28). In fact, the ordered Te vacancies induce the delocalization of electrons around the Pt atoms to promote the adsorption of hydrogen atoms on the Pt sites. Moreover, we have obtained the density of states (DOS) near the Fermi level (E$_f$), primarily stemming from the d orbitals, to probe the electronic interaction between the adsorbate and pristine PtTe$_2$ or other PtTe$_{2-x}$. The coupling between the adsorbate valence electron and the transition-metal d orbitals can result in the formation of split bonding and antibonding states.[71] The relative electron filling of the antibonding states depends on these energy levels relative to $E_f$, contributing to the bond strength, whereas the bonding states are often fully filled since they are far below $E_f$. The d-band center model is a useful approach for describing the adsorbate-metal interactions as the antibonding states are usually higher in energy than the d states.[70,72,73] The d-band centers ($E_d$) of $PtTe_2$, (2×2)- and (7×7)-$PtTe_{2-x}$ relative to $E_f$ are calculated to be -3.27, -2.97, and -3.11 eV, respectively (Figure S29), illustrating that $E_d$ moves closer to $E_f$ after introducing ordered Te vacancies. As a result, the upward shift of $E_d$, causing fewer electrons filling in the antibonding states, enhances the interaction between the adsorbate and surface, thereby boosting the Volmer adsorption process for (2×2)- and (7×7)-$PtTe_{2-x}$ catalysts compared to $PtTe_2$. Nonetheless, it is important to note that $E_d$ of (2×2)-$PtTe_{2-x}$ is closer to $E_f$ than that of (7×7)-$PtTe_{2-x}$, leading to stronger binding between the undercoordinated Pt sites and hydrogen atoms in (2×2)-$PtTe_{2-x}$, which in turn may hinder hydrogen desorption. Overall, ordered Te vacancies can facilitate electron delocalization around Pt atoms and elevate the d-band center, thereby activating the Pt sites in the basal plane for HER catalysis.[70]

In addition to the catalytic properties, we have also delved into the potential topological characteristics of the two $PtTe_{2-x}$ structures, given the topological nature of bulk $PtTe_2$[31,74] and (2×2)-$PtTe_{2-x}$.[33] The calculated band structures of (2×2)-$PtTe_{2-x}$ and (7×7)-$PtTe_{2-x}$ monolayers (Figure S30), both without and with spin-orbit coupling (SOC), indicate that both systems are metallic and exhibit visible Rashba band splitting due to the breaking of inversion symmetry by the Te vacancies. To determine the topological properties, we evaluate the topological invariant $Z_2$ using the Wannier charge center method.[75] It is noted that even though the systems have no global bandgap, there are well-defined curved chemical potentials (Figure S30), which can be utilized to identify the band topology.[76] The calculated $Z_2$ invariants summarized in Figure S30 indicates that both systems are topologically nontrivial with $Z_2 = 1$. Specifically, for (2×2)-$PtTe_{2-x}$, $Z_2 = 1$ around $E_f$, consistent with a previous study,[33] is further substantiated by the presence of Dirac-type edge states near 0.1 eV (Figure S31). For (7×7)-$PtTe_{2-x}$, the presence of two pairs of Dirac-type topological edge states (TESs) at ~-0.25 and 0.16 eV (Figure 5f,g and Figure S32) also confirms its 2D topologically nontrivial nature.

Given these distinct edge features and experimental observation of pronounced edge states (Figure S33), we have further investigated the hydrogen adsorption on different edges of (7×7)-PtTe$_{2-x}$. Our calculations reveal that the Pt sites and certain Te sites at the edges are also catalytically active for HER (Figure S34), potentially leveraging the topological nature of the edges to enhance the overall activity of the samples.[77] Therefore, it is expected that the vertically oriented PtTe$_{2-x}$ nanosheets, featuring abundant Pt and Te active edge centers, can further help these systems to serve as promising candidates for electrocatalytic HER.[24]

Normally, the catalytic activity of metal atoms in SACs is primarily dictated by the number, chemical identity, and arrangement of the nearest neighboring atoms, as well as their extended local environment.[78] Ensuring uniform catalytic activity across all metal-atom sites in SACs poses a challenge, given the difficulty in achieving identical atomic architectures around each metal atom.[12] In strong contrast, PtTe$_{2-x}$ superstructures not only achieve a comparable density of metal-atom sites to SACs but also expose periodically well-defined metal-atom sites with large-area uniformity. The distinct arrangements of Te vacancies in PtTe$_{2-x}$ superstructures result in varied hydrogen adsorption strengths, impacting HER performance. The (7×7)-PtTe$_{2-x}$ superstructure features eight types of undercoordinated Pt sites, each demonstrating favorable hydrogen adsorption strength for HER. Conversely, the (2×2)-PtTe$_{2-x}$ superstructure has only one type of undercoordinated Pt site with stronger hydrogen adsorption, which is less optimal for HER. Additionally, despite the slightly higher density of undercoordinated Pt sites on the (2×2)-PtTe$_{2-x}$ surface, the ECSA of the (7×7)-PtTe$_{2-x}$ surface is larger. This is possibly due to the following reasons: (i) The (2×2)-PtTe$_{2-x}$ surface may not be entirely covered by the (2×2) superstructure, containing regions of inert basal planes (Figure S2). (ii) The presence of small nanometer-sized islands on the (7×7)-PtTe$_{2-x}$ surface contributes to exposing abundant edge structures, thereby facilitating HER (Figure S5g). (iii) Certain Te sites at the edges of (7×7)-PtTe$_{2-x}$ are energetically favorable for hydrogen evolution. For 2D topological materials, TESs exist only at edges, potentially influencing the hydrogen adsorption Gibbs free energies of the atoms at edges. Considering SOC effect, the enhanced hydrogen adsorption strength of Pt sites at the edges of (7×7)-PtTe$_{2-x}$ (Figure S34e) may be due to the effective electron bath provided by TESs.[77,79] At present, it remains challenging to directly observe the contribution of topological properties to catalytic reactions. Nonetheless, these findings highlight the Te-vacancy lattices of PtTe$_{2-x}$ superstructures in enhancing HER performance through their well-defined and uniformly distributed catalytic sites.

## CONCLUSIONS

In summary, our work presents a significant advancement in the realm of electrocatalysis by introducing PtTe$_{2-x}$ superstructures with tailored properties. The (7×7)-PtTe$_{2-x}$ superstructure, in particular, emerges as a highly promising electrocatalyst for HER, combining excellent activity, exceptional stability, and topological characteristics. Moreover, the PtTe$_{2-x}$ structures feature flawless atomic quantum antidot lattices, suggesting potential applications in quantum information devices.[43,80] Nonetheless, the practical application of PtTe$_{2-x}$ nanosheets on a scalable level for applications in energy technologies necessitates methodological innovation. This may be achievable through the continuous processes involving the pre-introduction of excess metal atoms during single crystal growth for the creation of Te vacancies, electrochemical exfoliation for nanosheet production, and subsequent annealing for Te-vacancy migration.[16,43] Our findings not only contribute to the development of efficient and sustainable energy conversion technologies but also open up exciting prospects for the exploration of topological materials in catalysis and beyond.

## ASSOCIATED CONTENT

### Supporting Information

The Supporting Information is available free of charge at https://pubs.acs.org/doi/xxx.

Sample preparation methods, STM/LEED/AES/XPS/ARPES characterization methods, TEM characterization methods, catalytic performance evaluation, DFT calculations, and additional sample characterizations (STM, LEED, AES, XPS, ARPES, HRTEM, etc.) are supplied as Supporting Information.

## AUTHOR INFORMATION

### Corresponding Author

(Y.M.) Email: mustc@henu.edu.cn
(F.M.) Email: mingff@mail.sysu.edu.cn
(P.C.) Email: cuipg@ustc.edu.cn
(X.Xiao) Email: xdxiao@whu.edu.cn

### Author contributions

X.Xu, X.W. and S.Y. contributed equally to the work. Y.M., F.M., and X.Xiao supervised the project. X.Xu, S.Y., Y.M., C.W., H.L., and J.Y. performed all the experiments. X.Xu, Y.M., X.W., and G.L. conducted data analysis. X.W., G.L., K.W., P.C., and Z.Z. performed the DFT calculations. Y.Z. prepared the cross-sectional samples. L.L. and Y.M. performed TEM characterizations and data analysis. All other authors contributed to the scientific discussions. The manuscript was written by X.Xiao, Y.M., F.M., and X.Xu with the contributions from all coauthors.

**Notes**

The authors declare no competing financial interest.

**ACKNOWLEDGMENTS**

F.M. acknowledges the Guangdong Basic and Applied Basic Research Foundation (2020B1515020009), National Science Foundation of China (12174456) and the Fundamental Research Funds for the Central Universities, Sun Yat-sen University (22qntd0503). Y.M. acknowledges the support from Natural Science Foundation of Henan Province (242300420627) and Start-up Fund of Henan University (CX3050A0970531). X.Xiao acknowledges the financial support from Start-up Fund of Wuhan University (1302/600460055). P.C. and Z.Z. acknowledge the National Science Foundation of China (11974323 and 12374458) and Innovation Program for Quantum Science and Technology (2021ZD0302800). We thank the Core Facility of Wuhan University for their assistance with Raman and TEM measurements. Y.M. thanks Dr. Peng He and Dr. Qiulan Zhou for helpful discussions.

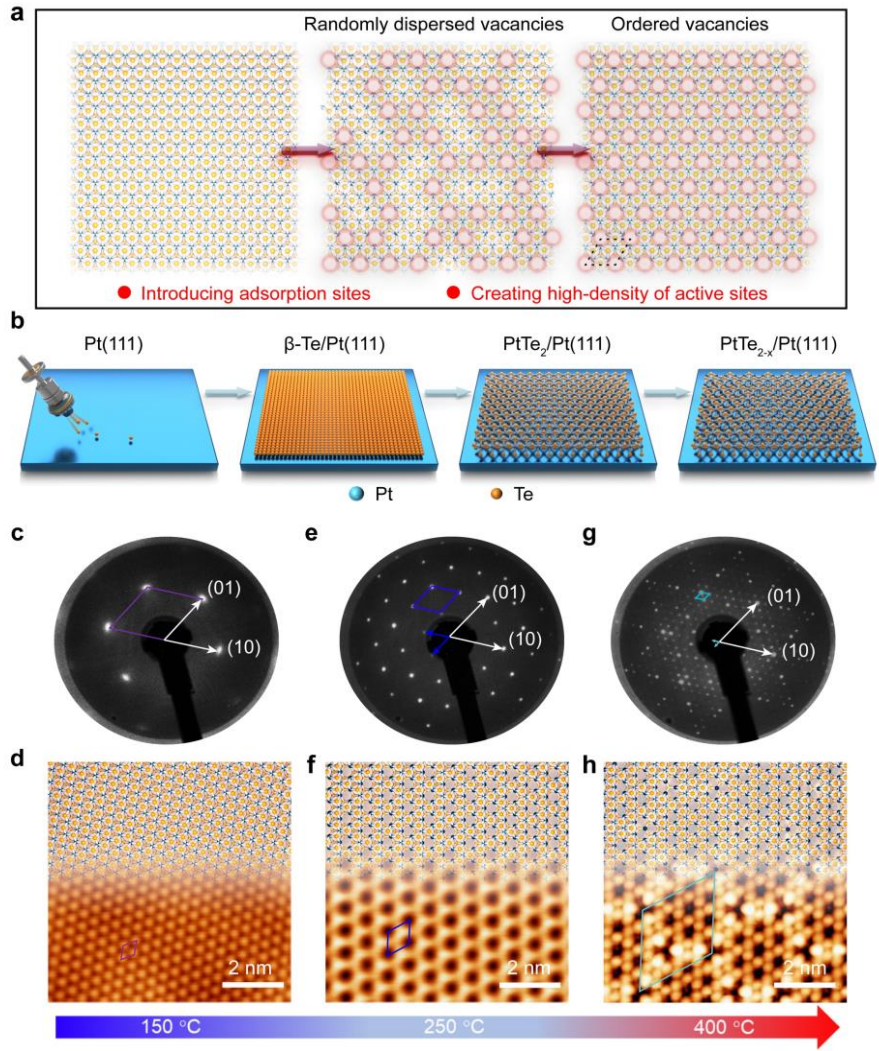

**Figure 1.** Sample fabrication strategy. (a) Scheme to expose active sites for improving the catalytic performance of TMDs. (b) Strategy for the preparation of platinum telluride superstructures. (c,e,g) LEED patterns of various platinum telluride samples grown at different annealing temperatures: (c) PtTe$_2$ at 150 °C, (e) (2×2)-PtTe$_{2-x}$ at 250 °C, and (g) (7×7)-PtTe$_{2-x}$ at 400 °C. All the LEED patterns are obtained with the electron energy of 100 eV. (d,f,h) High-resolution STM images of (d) PtTe$_2$ ($V_b$ = 0.5 V, $I_t$ = 1 nA), (f) (2×2)-PtTe$_{2-x}$ ($V_b$ = 0.5 V, $I_t$ = 2 nA), and (h) (7×7)-PtTe$_{2-x}$ ($V_b$ = 1.5 V, $I_t$ = 0.1 nA). The corresponding atomically structural models are also superimposed. The orange and blue balls represent the Te and Pt atoms, respectively.

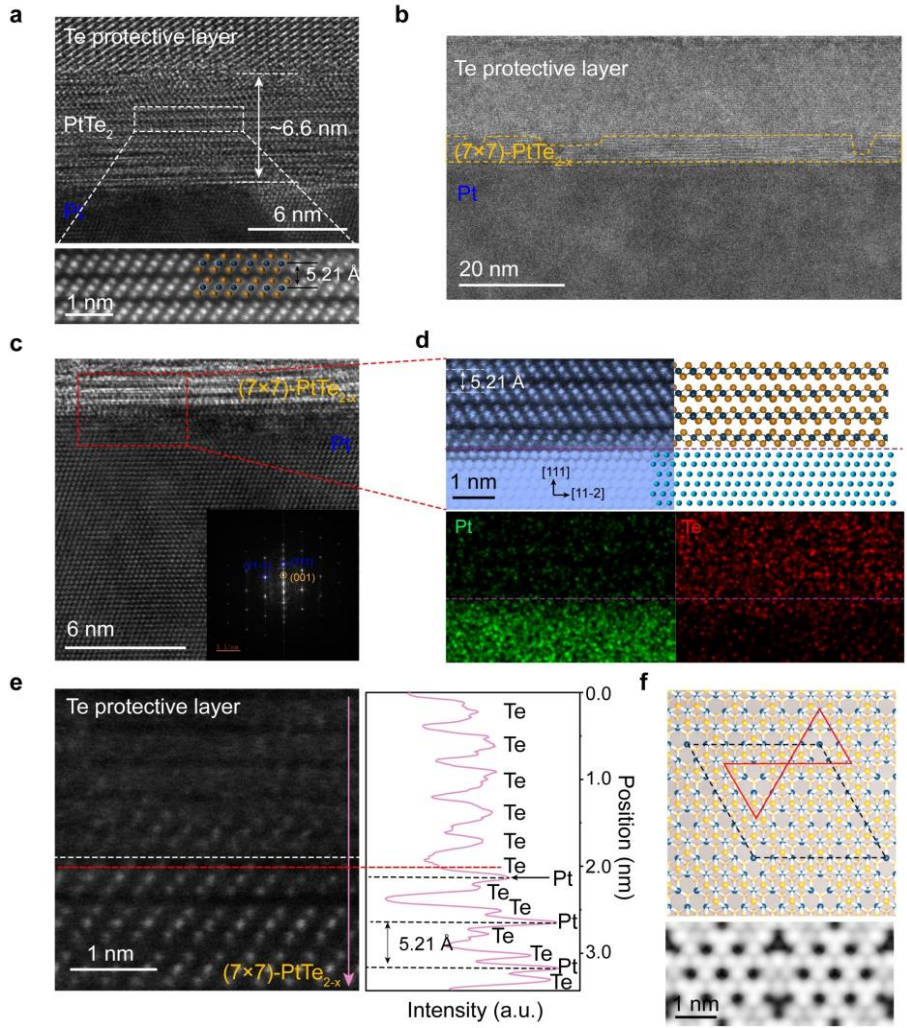

**Figure 2.** Atomic structure of (7×7)-PtTe$_{2-x}$. (a) Large-scale side-view TEM image (upper) and atomic-resolution HAADF-STEM (bottom) of PtTe$_2$ sample. Blue and orange balls represent Pt and Te atoms, respectively. To mitigate ion beam damage during focused ion beam milling, a Te protective layer was deposited at room temperature prior to the milling process. (b) Large-scale side-view TEM image of (7×7)-PtTe$_{2-x}$/Pt(111). (c) High-resolution TEM image of cross-sectional quadra-layer (7×7)-PtTe$_{2-x}$/Pt(111). The inset shows the corresponding fast Fourier transform (FFT) image. (d) Atomic-resolution HAADF-STEM image near the (7×7)-PtTe$_{2-x}$/Pt(111) interface with the structural model and corresponding element mapping images. (e) Atomic-resolution HAADF-STEM image near the (7×7)-PtTe$_{2-x}$ surface and corresponding intensity profile along the direction of the image. White and red dash lines represent the Te/(7×7)-PtTe$_{2-x}$ interface and the vertical position for Te top layer of (7×7)-PtTe$_{2-x}$ surface layer, respectively. (f) Structural model for (7×7)-PtTe$_{2-x}$. Upper image shows the atomic structure with the unit cell indicated by the dashed black line. Two red triangles highlight the domain boundary. Bottom image shows simulated STM image.

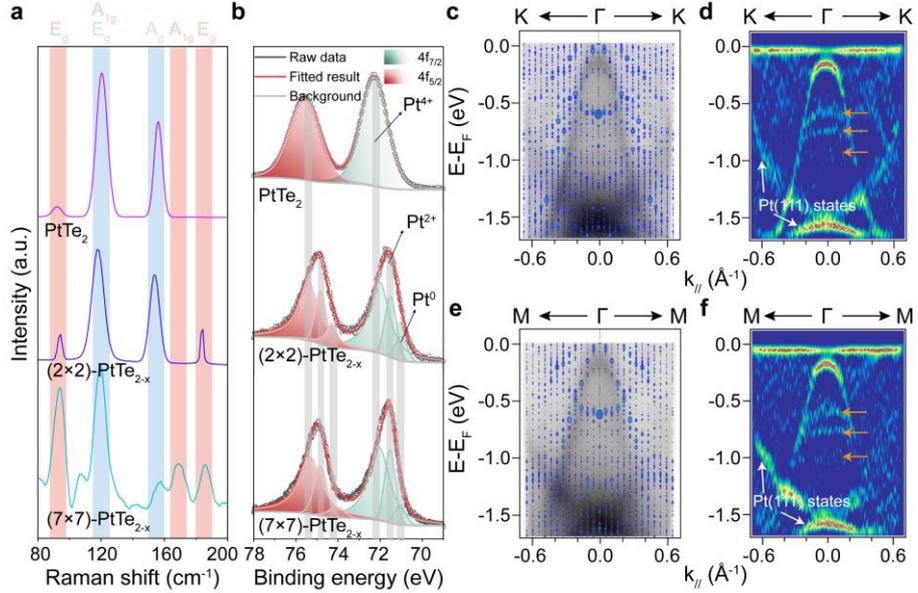

**Figure 3.** Characteristics of (7×7)-PtTe$_{2-x}$ film on Pt(111). (a) Raman spectra of PtTe$_2$, (2×2)-PtTe$_{2-x}$, and (7×7)-PtTe$_{2-x}$. (b) XPS spectra of Pt-4f core levels for PtTe$_2$, (2×2)-PtTe$_{2-x}$, and (7×7)-PtTe$_{2-x}$. (c,e) ARPES maps of (7×7)-PtTe$_{2-x}$ along the (c) ΓK and (e) ΓM directions taken at ~7 K, superimposed with the calculated band structures. (d,f) Corresponding maps of the second-order derivative of the ARPES intensity with respect to the energy to (c) and (e), respectively. The signals near the Fermi level are artifacts, caused by the second-order derivative process.

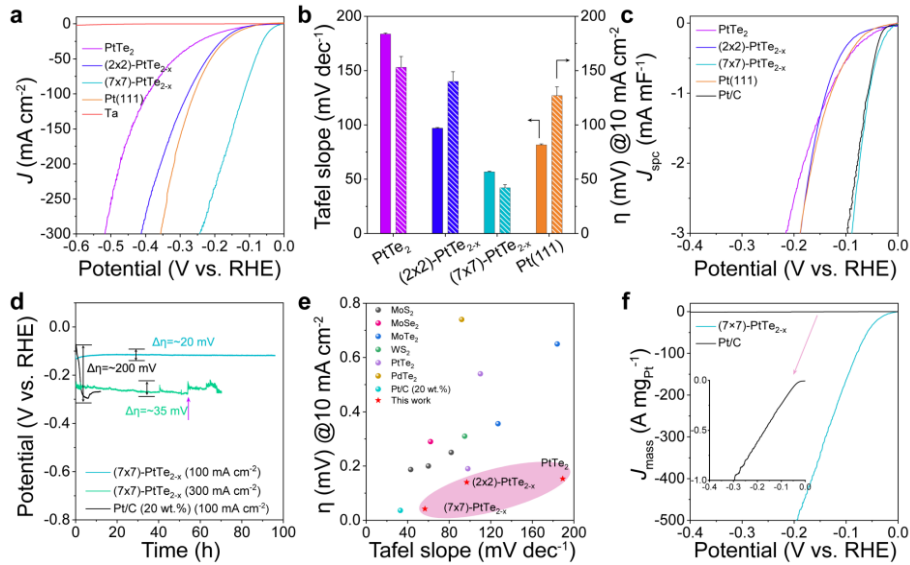

**Figure 4.** HER performance and operation stability of PtTe$_{2-x}$ in an acidic electrolyte. (a) LSV curves of various Pt-based samples and Ta substrate in 0.5 M H$_2$SO$_4$ at a scan rate of 5 mV s$^{-1}$ with 80% iR compensation. (b) The corresponding Tafel slopes and overpotentials at 10 mA cm$^{-2}$. (c) Corresponding specific HER activities of Pt-based catalysts. (d) Chronopotentiometry test of the (7×7)-PtTe$_{2-x}$ sample recorded at 100 mA cm$^{-2}$ for over 90 h and at 300 mA cm$^{-2}$ for over 70 h. The purple arrow indicates the time for replacing with fresh 0.5 M H$_2$SO$_4$ solution. (e) Performance comparison of (2×2)- and (7×7)-PtTe$_{2-x}$ with other reported TMD-based HER catalysts. (f) Comparison of mass activities of the (7×7)-PtTe$_{2-x}$ and Pt/C catalysts.

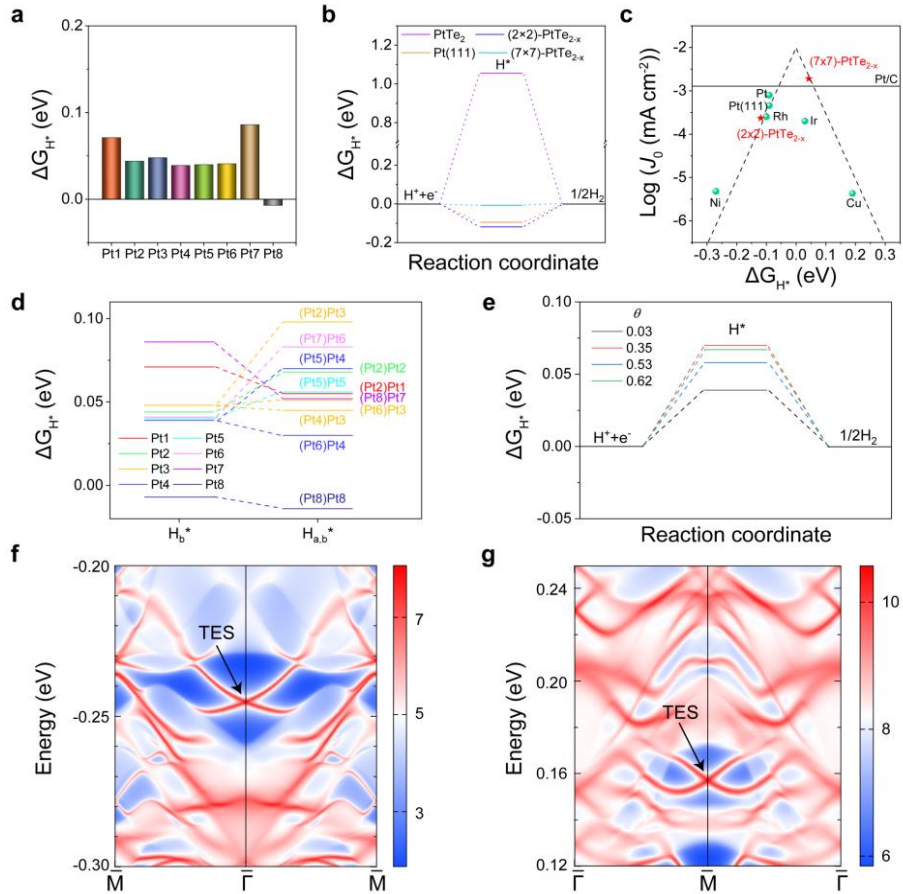

**Figure 5.** Binding strength and topological edge states. (a) Hydrogen adsorption Gibbs free energies ($\Delta G_{H^*}$) on different undercoordinated Pt sites in the (7×7)-PtTe$_{2-x}$ structure. The labels of the undercoordinated Pt sites refer to Figure S26. (b) Hydrogen adsorption free energy diagram on various Pt-based samples. (c) Volcano plot of the PtTe$_{2-x}$ samples for the relationship between the exchange current density ($J_0$) and hydrogen adsorption Gibbs free energy. The Gibbs free energy of the (7×7)-PtTe$_{2-x}$ superstructure is denoted as the weighted average value of the Gibbs free energies associated with all types of undercoordinated Pt sites. The cases of Pt(111), Rh, Pt, Ir, Ni, and Cu are included for comparison.[66] (d) Free energy diagram for hydrogen adsorption on two nearest neighboring sites among Pt1 to Pt8 in the (7×7)-PtTe$_{2-x}$ superstructure. H$_b^*$ and H$_{a,b}^*$ represent the states of one hydrogen adsorbed on the Pt$_b$ site and two hydrogen adsorbed on two nearest neighboring Pt$_a$ and Pt$_b$ sites, respectively. (e) Free energy diagram for hydrogen adsorption on (7×7)-PtTe$_{2-x}$ at different hydrogen coverages ($\theta$). Here, a surface unit cell of (7×7)-PtTe$_{2-x}$ was adopted, and the total number of undercoordinated Pt sites on the surface is 34. (f,g) Topological edge states with the Dirac nature shown at high-symmetric points located around (f) -0.25 and (g) 0.16 eV. The warmer colors denote higher local density of states, and the blue regions denote the bulk bandgaps.

For table of contents only

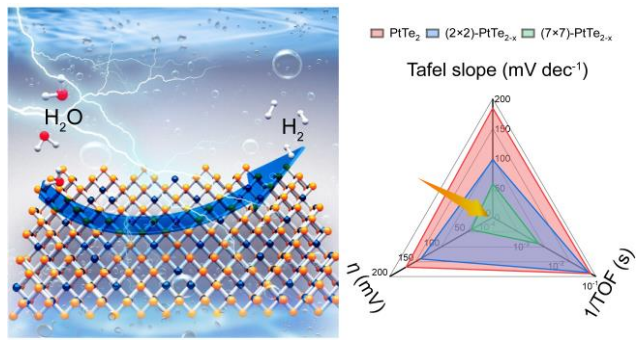